\documentclass[aps,floatfix,nofootinbib,superscriptaddress,11pt]{revtex4}

\usepackage{tikz}
\usetikzlibrary{calc,decorations.markings}

\usepackage[T1]{fontenc}
\usepackage{color}
\usepackage[latin9]{inputenc}
\usepackage{verbatim}
 \usepackage{mathtools}
 \usepackage{tcolorbox}
 \usepackage{float}

 \usepackage{tikzfeynman}

\usepackage{textcomp}
\usepackage{appendix}
\usepackage{amstext}
\usepackage{amssymb}
\usepackage{graphicx}
\usepackage{setspace}
\usepackage{hyperref}

\usepackage{graphicx,bm,psfrag, amsmath, bbm}

\newcommand{\beq}{\begin{equation}}
\newcommand{\eeq}{\end{equation}}
\newcommand{\beqa}{\begin{eqnarray}}
\newcommand{\eeqa}{\end{eqnarray}}
\newcommand{\ba}{\begin{array}}
\newcommand{\ea}{\end{array}}

\usepackage{youngtab}

\newcommand{\be}{\begin{equation}}
\newcommand{\ee}{\end{equation}}
\newcommand{\bea}{\begin{eqnarray}}
\newcommand{\eea}{\end{eqnarray}}

\newcommand{\dd}{\mathop{}\!\mathrm{d}}

\newcommand{\intp}[1]{\int \frac{\dd^d #1}{(2\pi)^d}}

\definecolor{rossoCP3}{cmyk}{0,.88,.77,.40}
\definecolor{blue1}{rgb}{0.2,0.2,0.6}
\definecolor{light-gray}{gray}{0.85}

\hypersetup{pdfstartview=FitV,colorlinks=true,linkcolor=blue,citecolor=red,filecolor=black,urlcolor=blue}

\begin{document}

\title{\boldmath UV Completion on the Worldline}

\author{ Steven Abel}
\email{\tt s.a.abel@durham.ac.uk}
\affiliation{\mbox{IPPP,
Durham University, South Road, Durham, DH1 3LE}}
\author{Nicola Andrea Dondi}
\email{\tt dondi@cp3.sdu.dk}
\affiliation{\mbox{IPPP,
Durham University, South Road, Durham, DH1 3LE}}
\affiliation{{\color{rossoCP3}CP${}^3$-Origins} \& the Danish Institute for Advanced Study,
Univ. of Southern Denmark, Campusvej 55, DK-5230 Odense}

\date{\today}

\begin{abstract}
\vskip2cm
\noindent A framework for UV completing particle theories is proposed, based on the worldline formalism, which is equivalent to weighting all sums over histories with a proper-time dependent measure that has a smallest proper-time. We deduce a criterion to avoid ghosts, 
and find that the ghost-free theories have the (string-theory inspired) infinite-derivative model of Siegel as a special case, but represent a significant generalisation. 
Generically, the UV of such theories is dominated by a saddle-point in the proper-time, leading to simplification in the  computation of amplitudes. 
We focus on a particularly attractive option which is to mimic the regulating properties of modular invariance with a worldline  ``inversion symmetry''.

\vskip8.7cm
{\noindent \footnotesize Preprint:  CP3-Origins-2019-19 DNRF90,    
IPPP/19/36}

\end{abstract}
\maketitle
\newpage


\section{Overview and relation to infinite derivative field theory}

The finiteness of string-theory can be attributed to two features. First -- focussing on closed strings --  due to  
modular symmetry the ultra-violet  (UV) region of the integral over the 
modular parameter is excised from the fundamental domain of one-loop diagrams. Similar excisions occur 
at higher loops. Therefore it seems  likely
that in a fully consistent theory modular invariance forbids UV divergences to all orders. A somewhat different perspective 
is provided by the behaviour of the worldsheet Green's functions at short distances. Even at tree-level 
they yield amplitudes that are exponentially suppressed at high 
momentum (see for example \cite{grossmende,Siegel:2003vt}). Exponential suppression can also be seen 
in string field theory propagators which are dressed by factors of $e^{-\square /M_s^2} $ where $M_s$ is the fundamental scale, 
as exemplified in \cite{stringfield,Tseytlin:1995uq,Siegel:2003vt}. 
Equivalently by a field redefinition such suppression can be attributed to the cubic string field interactions (see for example \cite{Calcagni:2007ef}). The general conclusion is that 
amplitudes appear to attenuate exponentially in the UV above the scale $M_s$. 

Motivated by these properties of string-theory, we wish to propose a 
particle framework that has the same benefits, built from the ground up. Our approach is to work within the worldline formalism  \cite{feyn, Affleck:1981bma, Bern:1991aq, Strassler:1992zr,
Schmidt:1993rk,Schmidt:1994zj}, 
which allows us to mimic closely the behaviour of first quantised strings. (For reviews of the worldline formalism see \cite{Schubert:2001he}.) 

As a starting point, consider the Schwinger parameterised scalar particle propagator  (suitably Euclideanised),
\be
\Delta(p^2)~=~ \frac{1}{p^2+m^2} ~=~ \int_0^\infty \dd T  e^{- T (p^2+m^2)} ~,
\label{eq:StandardProp}
\ee
where $T$ is the real Schwinger proper-time. Propagators naturally appear in this form 
in the ``particle limit'' of string-theory. For example, in a closed theory the role of $T$ is played by the imaginary 
part, usually denoted $\tau_2$, of the modular parameter 
in the infra-red (IR) where it is large. However modular invariance dictates that, as one approaches the UV cusp where the modular parameter vanishes,  
the exponent instead goes like $1/\tau_2$ (since this region can always be mapped back to large $\tau_2$ by a $\tau_2\rightarrow 1/\tau_2 $ M\"obius transformation). 
It is tempting to copy this behaviour in the particle context, by modifying the 
propagator so that it is written as an integral over the single real parameter $t$ as follows:
\be
\label{master}
\Delta(p^2) ~=~ \int_0^\infty \dd t \, e^{- T(t) (p^2+m^2)} ~,
\ee
where the proper-time is some function of $t$ that reproduces the correct IR behaviour, but also has a worldline ``inversion symmetry'', corresponding to the only surviving M\"obius transformation
\footnote{Note that as we integrate over the whole of $t$ (which has only 2 copies of the fundamental domain) we do not need to make the measure invariant.}:  
\begin{align}
\label{limits}
{\rm a})\qquad \lim_{t\rightarrow \infty}\frac{T(t)}{t} &~=~  1 ~, \nonumber \\
{\rm b})\, ~\qquad\qquad T(t) &~=~  T(t^{-1} ) ~.
\end{align}Let us consider the simplest option\footnote{Despite the superficial similarity this duality is {\it not} the same as the one described in \cite{Tduality} (see also \cite{Hossenfelder:2012jw} for a review). That duality is a space-time one equivalent to $T\rightarrow 1/T$ and there is no lower bound on $T$. (The distinction is the same as coordinate-space duality versus modular invariance).}. 
\be
\label{simplet}
T ~=~ t+t^{-1}~,
\ee 
where we henceforth choose units in which the fundamental scale is one. 
Performing the Schwinger integral we find the propagator to be 
\be
\Delta(p^2) ~=~ 2 K_1( 2 (p^2+m^2))~,
\ee
where $K_1$ is the modified Bessel function of the second kind. It has the following asymptotic behaviour:
\be
\label{delta}
\Delta(p^2) ~\longrightarrow ~
\begin{cases*}
                    ~ \frac{1}{p^2+m^2}   & ; ~ $p^2 \ll 1 $~,  \\
                     ~ \frac{\sqrt{\pi} e^{-2(p^2+m^2) } }{\sqrt{p^2+m^2}} & ; ~$p^2 \gg 1$~.
                 \end{cases*} %
\ee
As well as exhibiting desirable exponential 
suppression at momenta above the fundamental scale, $\Delta(p^2)$ has the interesting 
property that it possesses only the single physical pole near the origin of $z=p^2+m^2$. Indeed 
$ K_1(z)$ is holomorphic in the right-half complex plane. Otherwise it has only a branch-cut for the higher derivative terms along the negative real axis emanating from the physical pole at $z=0$, so a theory with such a propagator can be considered to be ghost-free. 
The modification in  \eqref{master} can thus be thought of as a means of generating an infinite-derivative, ghost-free and finite 
field theory similar to (but more general than) those in refs.\cite{Siegel:2003vt,Biswas:2005qr,Biswas:2011ar,Buoninfante:2018mre}. 
(By ghost-free here we mean that there are no additional poles with negative norm, but as in those theories there are actually no additional poles at all.)

 Indeed, dropping the second inversion-symmetry requirement in (\ref{limits}\textcolor{blue}{b}), it is straightforward to show that: 
 \begin{quote}
 {\it Any $T(t)$ for which Re$(T)>0$ for all $~t>0$, and  
$t T(t^{-1})$ is entire generates a ghost-free infinite-derivative theory.} 
\end{quote}
The proof of this statement will
be given in the following section. Let us for the moment assume the statement to be true, and consider its implications and physical  interpretation. 
   
The trivial example of a function that obeys the above condition is $T(t)= t+1 $, which gives the exponentially suppressed 
propagator, $\Delta(p^2) = e^{-(p^2 + m^2) } /(p^2+m^2)$, advocated in \cite{Siegel:2003vt,Biswas:2005qr,Biswas:2011ar,Buoninfante:2018mre}. This case is effectively a lower cut-off on $T$ (as also pointed out in \cite{Sen:2016gqt}) and is in fact the only situation in which our prescription 
equates precisely to that of \cite{Siegel:2003vt,Biswas:2005qr,Biswas:2011ar,Buoninfante:2018mre}, in the sense that the resulting propagator is 
the usual field-theoretic one multiplied by a form-factor which is an entire function of $p^2$.
However there are an infinite number of ghost-free theories that can be 
obtained this way\footnote{We should add that in effect the full string propagator {\it does} contain an infinite number of physical poles \cite{Cohen:1985sm}, which the 
exponentially suppressed single pole version approximates, via the Stirling formula.}. 
It is also easy to see that eq.\eqref{simplet} (plus a constant) is the unique choice that is $t\rightarrow 1/t $ symmetric\footnote{In the symmetric case both $z T(z)$ and $z T(z^{-1})$ are entire. Hence,  $z T(z)$ can be represented as a power series, whose highest power is $z^2$ due to the large $z$ limit in (\ref{limits}a). Likewise, considering the lowest power of $z T(z^{-1})$ in the small $z$ limit, its lowest power is $z^0$.}. 

Conversely, the 
{propagators generated by our procedure can be seen as coming from general infinite-derivative actions of the form
\begin{equation}
S ~=~ \int \dd^d x\,\,\, \phi \, \Delta(-\square)^{-1} \phi~.
\end{equation}}
For example, in the 
$t\rightarrow 1/t$ symmetric case the Euclidean propagator can be rewritten as $\Delta(p^2) = F(p^2+m^2) \frac{1}{p^2+m^2}$ where the form-factor is 
\begin{align}
 F (z) ~=~ 2 z K_1( 2 z) ~.
\end{align} 
The crucial distinction between  our procedure and that of  \cite{Siegel:2003vt,Biswas:2005qr,Biswas:2011ar,Buoninfante:2018mre} is therefore that 
it generates ghost-free theories in which  the 
function $F(z)$ is related to a simple worldline prescription but need not be entire: like the example above it may just be holomorphic in the right-half plane, with possible branch-cuts in the left-half plane at higher order in $z$. 
Like the theories of  \cite{Siegel:2003vt,Biswas:2005qr,Biswas:2011ar,Buoninfante:2018mre}, the form-factor tends to one at small momenta, and decays exponentially at large momenta, and like those theories 
we expect these more general ones to be both non-local and acausal (which we revisit below), but only on scales shorter than the fundamental scale. 

What is the meaning of the proper-time redefinition?  It is of course always possible to make a substitution to bring the propagator back to its original form as an integral over $T$, whereupon we find an interpretation in terms of minimum proper-time: 
\begin{align}
\label{reparammed}
\Delta(p^2)~&=~ \int_{T_0} ^\infty \dd T \mbox{$ \left(\frac{1}{T_+'} -  \frac{1}{T_-'}\right)$} \ e^{- T (p^2+m^2)} ~,\nonumber \\
~ & =~ \int_{2} ^\infty \dd T 
\mbox{$ \frac{T}{\sqrt{T^2-4} } $}  ~   e^{- T (p^2+m^2)} ~,
\end{align}
where in the first line we add the two branches $T_\pm $ corresponding to $ t\in (0,1)$ and $ t \in (1,\infty)$ respectively (with $T'_\pm \equiv dT_\pm / dt$), and where the second line is 
specific to the example of \eqref{simplet}. In other words, our prescription 
  is equivalent to introducing
a weighting on any sum over histories which tends to one at large $T$, and diverges
at some (cut-off) proper-time, but 
slowly enough so as to leave a finite path integral (as an inverse square-root in this case). 

To support this interpretation, we can take the Fourier transform to obtain the propagator in target-space:  
\begin{align}
\label{xprop}
\Delta(x,y) &~=~ \intp{p} e^{-ip(x-y)} \int_0^{\infty} \dd t \, e^{- T(t) (p^2 + m^2)} \nonumber \\
&~=~  \int_0^{\infty} \dd t \, \frac{1}{(4\pi T)^{d/2}} e^{- \left[ \frac{(x-y)^2}{4T} + T m^2 \right]}~.
\end{align}
Considering $m=0$ for example, the Euclidean picture is of an integral over solutions to the diffusion equation in $d=4$ dimensions.
Broadly speaking, in a path integral the initial data in coordinate-space is sampled with Gaussians that diffuse outwards with the proper-time of the path.  
However a minimal value for $T$ means that the $\delta$-function at $T=0$ is no longer available, so the sampling is always smeared by
at least the fundamental scale: as in string-theory, physics has now acquired a minimum length\footnote{ 
Our prescription is also equivalent to adopting a (proper) time dependent 
diffusion coefficient, $D(t)=(1-1/t^2)$.  It may be  interesting to speculate on the fact that such time-dependent coefficients can be realised in colloidal and random-walk systems.}.

Finally,  we would like to consider perturbation theory in the worldline formalism, and for this we need to interpret  $T(t)$ in the context of a worldline theory. Following the standard treatment of the point particle (see for example \cite{Cohen:1985sm}) we wish to rewrite the propagator in a manifestly reparameterization invariant way by introducing a worldline parameter $\tau$ with an einbein $e(\tau)$: in ordinary field theory one has
\begin{equation}
\Delta(x,y) ~=~ \int_{x(0)=x}^{x(1)=y} \frac{\mathcal{D}x^{\mu} \mathcal{D} e}{\text{Vol(Gauge)}} e^{- \int_0^1 \dd \tau \left[ \frac{\dot{x}^2}{2e} + \frac{e m^2}{2} \right] } ~,
\label{eq:PropRep}
\end{equation}
where the einbein functional measure is usually defined from the norm and functional measure in its own tangent space: 
\begin{equation}
||\delta e ||^2 ~=~ \int \dd \tau \, e^{-1} \delta e^2~, \quad \int \mathcal{D}(\delta e)\, e^{-\frac{1}{2} ||\delta e||^2} ~=~ 1~.
\label{eq:DefMeasure}
\end{equation}
With this definition one can show that $\mathcal{D} e = \mathcal{J}\, \dd T\, \mathcal{D}\tilde{e} $, with $T$ being the modulus (here identified with the worldline length, $T=\int^1_0 \dd \tau e$) and with $\tilde{e}$ parameterising the pure gauge part of $e$ (identified as $\tilde {e} = e - T$, such that $\int^1_0 \dd \tau \tilde{e}\,=\,0$). The jacobian $\mathcal{J}$ can be computed in $\zeta$-function regularisation and amounts to a constant. The integral \eqref{eq:PropRep} can be carried out and reproduces the standard propagator in eq.\eqref{eq:StandardProp}. 

The conventional exponentially suppression form-factor of \cite{Siegel:2003vt,Buoninfante:2018mre} corresponds to an alternative and equally consistent reparameterisation invariant path integral that can be derived from \eqref{eq:PropRep} by restricting the integral over $e$ to a diff-invariant domain:
\begin{equation}
\int \mathcal{D}e \rightarrow \int_{D} \mathcal{D}e, \quad \text{where} \quad D: \{ e(t)\, |\,\, \text{if} \,\, e \in D \,\, \text{then} \,\, f' e(f(t)) \in D \}~.
\end{equation}
This is satisfied by a simple lower bound on the modulus $T$ which, as mentioned above and evidently from \eqref{reparammed}, is equivalent to the special case in which $T$ is defined in terms of a worldline parameter $t$ as 
 $T(t)=t+1$. From the worldline field theory perspective this choice is neither more nor less consistent than the standard one. 

Extending this correspondence, one can instead define the integral over the einbein such that it reduces to a {\it weighted} integral over the modulus as in \eqref{reparammed}, while retaining reparametrisation invariance. 
%
This is possible by employing a $T$ dependent einbein norm,
\begin{equation}
|| \delta e ||_{f} ~=~ \frac{1}{f(T)^4} ||\delta e||~,
\label{eq:NewNorm}
\end{equation}
where $f \rightarrow 1$ as $ T \rightarrow \infty$ in order to recover the usual propagator in the IR.
Following the steps described in \cite{Cohen:1985sm}, this definition leads to the target space propagator
\begin{align}
D(y,x) ~=~ \text{const.} \intp{p} e^{ip(y-x)} \int_{a}^{\infty} \dd T\, f(T)\, e^{- T(p^2 + m^2)}~,
\end{align}
where we include a lower bound $a$ in the integral as discussed earlier, which by a substitution is then rendered in the form \eqref{xprop}. In our example, $a=2$  coincides with the first singularity of $f(T)$ encountered approaching from large $T$, so that the norm (\ref{eq:NewNorm}) never becomes degenerate. (Note that the einbein norm goes to zero at $a$ slower than $(T-a)^2$ to maintain a finite path integral.)

The remainder of this work discusses the implications of our prescription, focussing on the behaviour of amplitudes at high momentum. After a brief derivation in the following section of the condition for ghost-freedom, we  present the general formalism for amplitudes, in particular the required vertex operators, for scalar QED. Extension to general gauge theories and to theories with fermions would follow in an obvious way from the existing worldline literature, so we will not include it in this paper. We then work through a succession of increasingly intricate diagrams, beginning at tree-level and passing on to multiple loops and legs. In many cases we will find significant simplification due to the dominant saddle at $t=1$. Finally we will argue that using a worldline prescription clarifies the procedure for passing to Minkowski-space, as it obviates the need to define an explicit Wick rotated propagator.

\section{The condition for ghost freedom}

Let us now turn to a proof of the condition for ghost freedom. We consider $\Delta(z)=\int_0^\infty \dd t e^{-T(t) z}$, in the Euclidean right-half $z$ plane. Condition (\ref{limits}\textcolor{blue}{a}) plus the entireness of $t T(t^{-1})$ implies that without loss of generality $T$ can be always expanded for all $t$ as  \begin{align} 
T &~=~ t+ \sum_{n=0} \frac{a_n}{t^n}~,
\end{align}
 for some generally complex coefficients $a_n$, since $t T(t^{-1})$ has  infinite radius of convergence. 
The constraint 
Re$(T)>0$ for all ~$t>0$ then implies that $\Delta(z)$ is finite and hence holomorphic everywhere in the right-half $z$ plane,
except for a physical pole which can appear when the exponent vanishes while $t \rightarrow\infty$. This occurs only at 
$z=0$. As the exponent is linear in $z$, this can only give a simple pole.

The finiteness and holomorphicity applies only in the right-half plane of $z$, because the Schwinger integral for $\Delta(z)$ diverges in the left-half plane (as it does normally of course). To consider the general 
analytic continuation to the left-half plane of $z$ 
we use standard techniques (see e.g. \cite{marino}). Consider $z=\rho e^{i\theta}$. Analytic continuation in $z$ is performed by  
counter-rotating the $t$ integration contour. There are generically two essential singularities when $t\rightarrow \infty $ and $t\rightarrow 0 $. We 
may treat them separately by splitting the $t$ integral into two pieces for $t<1$ and $t>1$. Taking the latter first, the contour for $t$ integration is 
counter-rotated by $e^{-i\theta }$ so that the integral becomes 
\be
\Delta(z) ~=~ e^{- i \theta } \int_1^\infty  dt  e^{- \rho t \, \frac{1}{t e^{-i\theta}}  T ( t e^{-i \theta}) }~.
\ee
The additional factor in the exponent $\frac{1}{t e^{-i\theta}}  T ( t e^{-i \theta}) $ is entire, so its large-$t$ limit is unity, regardless of $\theta$. Hence the integral is finite  and there are no poles for any $\theta$ except for the previous one  at $\rho=0$. However taking $\theta = \pm (\pi-\epsilon) $ generally reveals a discontinuity, and hence a branch-cut
along the negative real $z$ axis. A similar analysis can be performed for the $t<1$ part of the integral, by making the substitution $t\rightarrow 1/t$. \\

To check the above, we can consider the two special cases, of  $T=t+1$ and the Bessel function. In the first case the analytic continuation gives degenerate values for $\theta = \pm \pi$ so as expected there is no discontinuity. In the second case the integral can be evaluated at large $\rho$ by deforming to a steepest descent contour in $t$ going through the saddle at $t=e^{i\theta}$. The result when $\theta $ approaches $\pm \pi$ is $\Delta (\rho e^{i (\pi - \epsilon)}) -  \Delta (\rho e^{- i (\pi - \epsilon)}) \approx  - 2 i e^{2 \rho} \sqrt{\pi/\rho}$, which is the  asymptotic approximation to $4 \pi i I_1 (2 z) $ (i.e. the standard discontinuity for the $2K_1(2z)$).   

The presence of a branch-cut in the propagator is reminiscent of the situation in causal-set theory \cite{causal} (with the difference here being that as in \cite{Siegel:2003vt,Biswas:2005qr,Biswas:2011ar,Buoninfante:2018mre} we accept acausality on short scales). 
It is also similar to that in the large class of non-local theories discussed in  \cite{Barci:1995ad} (although there the theories have no simple pole part in the propagator). 

However it is also worth noting that for the example of the Bessel function, the branch-cut for small $|z|$ gives {\it higher} derivative terms in the propagator:
 \be 
 \Delta(z) ~=~ \frac{1}{z} + z \, (2 \gamma_E -1 + 2 \log \, z) + \ldots ~.
  \ee
 Therefore the effect of our modification on low energy physics is suppressed by factors of $p^2 /M_s^2$. More generally for a proper-time function  of the form $T=t+\frac{1}{t^m} + \ldots$ the first non-canonical term in $\Delta(z)$ is proportional to $z^{1/m}$, with the $T=t+1$ option giving a constant term proportional to $1/M_s^2$ to all propagators.

\section{Amplitudes at tree-level}

Evidently from the discussion of the previous section, the simplest case of an exponentially-suppressed propagator is indistinguishable from just putting a lower cut-off on the proper-time at  $T=1$. But as we shall now see, the advantage of the worldline prescription is that if
$T(t)$ 
has a minimum, as in the simple example of \eqref{simplet}, then many amplitudes become simple to evaluate, because they are dominated by a saddle-point.  

Let us begin by considering trees. The fact that the procedure can be understood as a 
weighting on the worldline integral, means that many results and techniques can be adopted wholesale (from e.g. \cite{Schubert:2001he}), with the 
modification arising only at the end of the calculation when one performs the integral over proper-time. 

Consider tree-level amplitudes in scalar QED. These can be obtained by covariantizing the momenta, and using a path integral 
representation of the scalar propagator, in which the gauge field $A_\mu$ appears as a Wilson line. In position space this gives
\begin{align}
\Delta(x,y)&~=~ \int_0^\infty \dd t e^{-T m^2 } \int_{x(0)=x}^{x(T)=y} {\cal D}x e^{-S[x,A_\mu]} ~,\nonumber \\
S[x,A_\mu] &~=~ \int_0^T \dd\tau  ~\frac{\dot{x}^2}{4} + i q \,\dot{x}\cdot A(x) ~,
\label{sss}
\end{align}
where $q$ is the charge of the scalar. 
From there one expands the gauge field as a sum of plane waves, 
\[ A_\mu(x(\tau) ) ~=~ \sum_{i=1}^n\varepsilon_{i,\mu}  e^{ik_i\cdot x }~,\] 
and extracts terms linear in all the polarization vectors. Passing back to momentum space one finds:
\begin{align}
{\cal A}^{(n)}  &~=~ (-iq)^n \delta^4(p_1+p_2 + {\scriptstyle \sum_i} k_i) \int_0^\infty {\dd t} \,e^{-  T (p_1^2+m^2) }\nonumber \\
&\qquad  \times
  \int_0^T \dd\tau_1\ldots \dd\tau_n  \,e^{(p_1-p_2)\cdot \sum_i (-\tau_ik_i-i \varepsilon_i) 
  }
e^{ (k_i \cdot k_j G_{ij} -2i \varepsilon_i\cdot k_j \dot G_{ij}+\varepsilon_i \cdot\varepsilon_j \ddot G_{ij} )}~,
  \label{An}
\end{align}  
where $G_{ij}=\frac{1}{2}|\tau_i-\tau_j|$ is the Green's function on the line, $p_1$ and $p_2$ are the momenta of the incoming and outgoing scalars, and one is instructed to extract the term in $\varepsilon_1\ldots \varepsilon_n$.   
Note that the length of the worldline $T$ appears as a modulus. The $\tau$'s parameterise the points of insertion on the worldline in the usual way, and the worldline Green's functions take the normal form in terms of these parameters.

The $n=1$ amplitude, for emission of a single photon (off-shell), gives 
\begin{align}
{\cal A}^{(1)} &~=~ -iq\,\delta^4(p_1+p_2 + k)\,\varepsilon \cdot (p_1-p_2) 
\int_0^\infty {\dd t} \,e^{-  T (p_1^2+m^2) }
\int^T_0 d\tau\, e^{-{\tau} (p_1-p_2) \cdot k },\nonumber \\
&~=~ iq\,\delta^4(p_1+p_2 + k)~\varepsilon \cdot (p_1-p_2) \,\frac{\Delta_{12}  }{p_1^2-p_2^2}~,
\end{align}
where  $\Delta_{12} = \Delta(p_1^2)  - \Delta(p_2^2)$. As anticipated, up to this point we have not needed to consider the details of the worldline prescription, however we can now insert the 
limits in \eqref{delta} to find (note that the external propagators have not yet been truncated)
\be
\label{delta12}
{ \frac{\Delta_{12}  }{p_1^2-p_2^2} }
\rightarrow 
\begin{cases*}
                     \frac{-1}{(p_1^2+m^2)(p_2^2+m^2)} & ; $p^2 \ll 1 $,  \\
                     {\small \frac{\sqrt{\pi} }{p_1^2-p_2^2}  \left(             \frac{e^{-2(p_1^2+m^2) } }{\sqrt{p_1^2+m^2}} -    \frac{e^{-2(p_2^2+m^2) } }{\sqrt{p_2^2+m^2}} \right)} & ; $p^2 \gg 1$,
                 \end{cases*} %
\ee
showing the expected exponential suppression.{ The impossibility of defining an amputated tree-level Green's function for $p^2 \gg 1$ reflects the non-locality of the theory: the interaction vertex cannot be resolved because it is not point-like.}

Note that double-photon emission at a point (a.k.a. the sea-gull) is included automatically in this prescription, as required by  gauge invariance. Explicitly, in the two photon case one brings down a $2\varepsilon_1 \cdot \varepsilon_2 \ddot{G}_{12} = \varepsilon_1\cdot \varepsilon_2 \delta (\tau_1-\tau_2) $ term  from the exponential in \eqref{An}. Integrating the delta function  over $\tau_2$ then leaves a single  $\varepsilon_1 \cdot \varepsilon_2 \,e^{(p_1^2-p_2^2)\tau_1}$ vertex to be integrated over the remaining single insertion position. (For more explicit details see the reviews in \cite{Schubert:2001he}.)

There is much less
restriction on how scalars are emitted, unless perhaps they are components of gauge multiplets (possibly higher dimensional or extended supersymmetric ones). In the case of charged scalars, an emission vertex must have pairs of bosons, so one can modify the action as 
\begin{align}
S[x,A_\mu,\phi] &~=~ \int_0^T \dd\tau  ~\frac{\dot{x}^2}{4} + i q \,\dot{x}\cdot A(x) + {\cal V}_{\phi\phi^*}(\phi(x))-m^2  ~,
\label{sss2}
\end{align}
where  ${\cal V}_{\phi\phi^*}(\phi(x))$ is the derivative of a potential of background scalars, ${\cal V}=m^2 |\phi|^2 + \frac{\lambda}{4} |\phi|^4 + \ldots $
In much of what follows we shall specialise to the case of ${\cal V}_{\phi\phi^*}(\phi(x))-m^2  \equiv \lambda |\phi|^2$. 

Adopting this case and expanding in plane waves  
\[ \phi (x(\tau) ) ~=~ \phi_0+ \sum_{i=1}^n  e^{ik_i\cdot x }~,\] 
we find an amplitude for $n$ scalar vertices (and hence the emissions of $2n$ scalars) which not surprisingly resembles 
the amplitudes for $2n$ photon emission with $n$ sea-gulls: 
\begin{align}
{\cal A}^{(n)}  &~=~ (-\lambda)^n \delta^4(p_1+p_2 + {\scriptstyle \sum_i^n} k_i) \int_0^\infty {\dd t} \,e^{-  T (p_1^2+m^2+\lambda |\phi_0|^2 ) }
  \int_0^T \dd\tau_1\ldots \dd\tau_n  \,e^{(p_1-p_2)\cdot \sum_i (-\tau_ik_i) 
  }
e^{ k_i \cdot k_j G_{ij} }~,
  \label{Anscalar}
\end{align}  
where now $k_i$ is the summed scalar momenta emitted from the $i$'th vertex. For example the single vertex amplitude (which is effectively the 4-point scalar coupling) is 
\begin{align}
{\cal A}^{(1)} 
&~=~ \lambda \,\delta^4(p_1+p_2 + k)~\,\frac{\Delta_{12}  }{p_1^2-p_2^2}~,
\end{align}
with the propagators written with effective mass-squareds $m_{\rm eff}^2 = m^2+\lambda |\phi_0|^2$. It experiences the same UV suppression {and vertex smearing} as the photon emission amplitude in \eqref{delta12}.

\section{One-loop amplitudes and threshold corrections} 

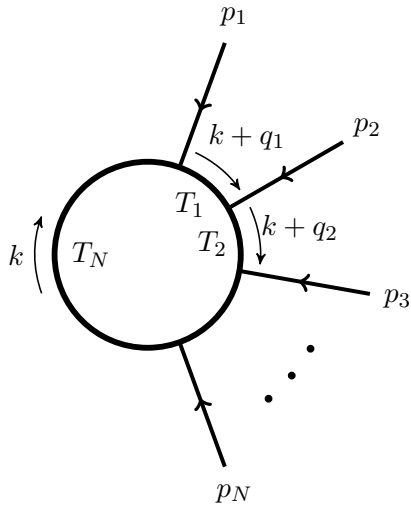
\begin{figure}[H]
\centering
	\begin{tikzpicture}[line width=1.5 pt, scale=2.5]	
		\draw[fill=black] (0,0) circle (.5cm);
		\draw[fill=white] (0,0) circle (.49cm);
		\draw[fermion] (70:1.2) -- (70:.5);
		\draw[fermion] (30:1.2) -- (30:.5);
		\draw[fermion] (-10:1.2) -- (-10:.5);
		\draw[fill=black] (-30:1) circle (.01);
		\draw[fill=black] (-40:1) circle (.01);
		\draw[fill=black] (-50:1) circle (.01);
		\draw[fermion] (-70:1.2) -- (-70:.5);
		\node at (70:1.35) {$p_1$};
		\node at (30:1.35) {$p_2$};
		\node at (-10:1.35) {$p_3$};
		\node at (-70:1.35) {$p_N$};
		\node at (180:0.7) {$k$};
		\node at (50:0.82) {$k+q_1$};
		\node at (10:0.82) {$k+q_2$};
		\node at (180:0.3) {$T_N$};
		\node at (50:0.35) {$T_1$};
		\node at (10:0.35) {$T_2$};
		\draw[->,>=stealth',semithick] (200:0.6) arc[radius=0.6, start angle=200, end angle=160];
		\draw[->,>=stealth',semithick] (65:0.6) arc[radius=0.6, start angle=65, end angle=35];
		\draw[->,>=stealth',semithick] (25:0.6) arc[radius=0.6, start angle=25, end angle=-5];
	 \end{tikzpicture}
	 \caption{One-loop $n$-point diagram, with the loop momenta and proper-time assignments.}
	 \label{fig:1loop}
	 \end{figure}

In order to pass to one-loop amplitudes, we need to be careful in adapting the general results outlined in \cite{Schubert:2001he}, because 
the expressions cannot now be resummed into a single logarithmic ``effective potential''. The 
$n$-vertex amplitude can be presented in a generically stringy form: 
\begin{align}
\label{ampscal}
\mathcal{A}^{(n)}_{1\ell} (\{p_i\})
 ~=~ &  
\int  \dd t  \, e^{-m^2   T(t)} \int_{S^1}{\cal D}x \, ~ V[p_1]...V[p_n] \, ~
e^{-S[x,A,\phi]}~,
\end{align}
where the action is as in \eqref{sss} and we omit symmetry factors. The path integral is over paths with $x(0)=x(T)$, and the $V$'s are vertex operators, taking the natural form 
\be
\label{vertex}
V_\lambda  [p]~=~ \lambda \int^T_0 \dd\tau  e^{ip\cdot x}~~; ~~~ 
V_A[p]~=~ \int^T_0 \dd\tau \varepsilon\cdot  \dot x \,e^{ip\cdot x}~,
\ee
for double-scalar and single-photon emission respectively. As in the tree-level case the momentum $p_i$ appearing in the former vertex operator is the sum of the two momenta emitted at that vertex. In order to implement the worldline procedure correctly, by manipulating the $\tau$ integrals the amplitude can be brought to a form that is democratic for the propagators. The result for the scalars (setting $A_\mu=0$) can be written as, see Fig. \ref{fig:1loop}:
\begin{align}
\mathcal{A}^{(n)}_{1\ell}(\{p_i\}) &= (-\lambda)^n \delta^d(\sum p_i ) \int  \frac{\prod_{i=1}^{n}  \dd t_i}{(\sum 4\pi T_i)^{d/2}} 
~ e^{-  \sum_i (q_i^2+m_{\rm eff}^2) T_i + \frac{\sum_{ij} q_i \cdot q_j T_{i} T_{j}}{\sum T_i} } + \text{perms.}
\label{qmaster}
\end{align}
where the $q_i$ are partial momentum sums:
\be
q_i ~=~ \sum_{j=1}^i p_j~,
\ee 
and where due to momentum conservation, $q_n=0$. 

This manipulation shows that the limits of $\{0,T\}$  on the integrals over vertex positions in \eqref{vertex} are correct, even though 
they may at first sight seems to violate the ``shortest distance'' paradigm. Indeed  
the $\tau$'s in \eqref{ampscal} correspond to rescaled Feynman parameters. For example, in the two-vertex case, 
$\tau$ corresponds to the usual Feynman parameter $u \equiv T_1/  (T_1+T_2)$, which still goes from $0$ to 1 despite the lower limit on $T_i$, with the identification $\tau \equiv T u$. Vertex operators are therefore still able to effectively coalesce when the overall size of the loop goes to infinity in the deep IR. We will now turn to several standard calculations to illustrate the effect on calculations in the worldline formalism. \\

\noindent \underline{\it Effective potential}:
It is interesting to determine the basic one-loop effective potential from $\mathcal{A}^{(n)}_{1\ell}(\{0\}) $ in the pure scalar theory. To evaluate the quadratic term (a.k.a. the Higgs mass correction) we note that the integral for $\mathcal{A}^{(1)}_{1\ell}$ is finite with no IR (i.e large $T$) divergences even if the state is massless. Reinstating $M_s$, setting $d=4$ and including a symmetry factor of $1/2$,  as $m_{\rm eff} \rightarrow 0$ the integral can be done explicitly. From this we infer the expected $M_s$ sized radiative contribution to scalar masses around $\phi=0$:
\be
{\cal V}^{(1)}_{\rm eff} ~\ni ~ \frac{1}{128 \pi}M_s^2 \,\lambda |\phi|^2 ~.
\ee
Note however that $m_{\rm eff}^2 $ grows as $\phi^2$, and at field values of $\phi \gtrsim M_s$ a saddle-point approximation around $t=1$ yields an exponentially suppressed potential of the form 
\be
{\cal V}^{(1)}_{\rm eff} ~ \longrightarrow  ~ \frac{\lambda |\phi|^2 M_s^2}{16 \pi^{\frac{3}{2}} (2+4\lambda |\phi |^2   )}   \,
e^{-2 |\phi|^2/M_s^2}~.
\ee 

For the quartic coupling we have 
\begin{equation}
{\cal V}_{\rm eff}^{(2)}  ~=~  -\frac{\lambda^2|\phi|^4 }{64\pi^2} \int \frac{ \dd t_1 \dd t_2}{(T_1 + T_2)^{2}} e^{-m_{\rm eff}^2(T_1 + T_2) }~.
\end{equation}
Again if $m_{\rm eff}$ is large one can use the saddle approximation which we will revisit later. However the interesting part is the 
fact that the integral also has a logarithmic term away from the saddle corresponding to the Coleman-Weinberg potential. 
To evaluate it, the reparameterized form of the integral in \eqref{reparammed} is useful. The integral gets contributions up to $T_1+T_2 \sim 1/m_{\rm eff}^2 $, so  in the limit that $m_{\rm eff}^2 \ll 1$, it becomes independent of the change in the measure in \eqref{reparammed} because it is one over most of the region of integration. The effective quartic interaction is then accurately approximated by 
\be 
{\cal V}_{\rm eff}^{(2)}  ~\approx~  \frac{ \lambda^2 |\phi |^4 }{64\pi^2} \log \left( 4 e^{\
\gamma_E+1-\frac{\sqrt{\pi}}{2}} \,\frac{ \lambda |\phi |^2 }{M_s^2 }\right) ~ .
\ee 
\noindent \underline{\it Threshold corrections}:
As in any UV completion, one can ``integrate out'' and match the full theory on to an effective field theory (EFT) 
with threshold corrections. In this case they encapsulate the difference between the standard integration over 
proper-time and the weighted one. The two-vertex calculation above can be adapted directly, as the relevant 
diagram is the vacuum polarisation diagram with scalars in the loop. We follow the procedure in \cite{Schubert:2001he} to extract the 
tensor pre-factor. In addition we can temporarily suspend momentum conservation, to express the result in terms of $s=(p_1+p_2)^2$:
\begin{align}
\label{thresh}
\mathcal{A}^{(2)}_{1\ell}(\{p_i\}) &~=~ 
(p^\mu p^\nu -g^{\mu\nu} p^2) 
\frac{ b_s}{(4\pi)^{d/2}} \int \frac{ \dd t_1 \dd t_2 } {(T_1+T_2)^{d/2}} 
 \times e^{- m^2 (T_1+T_2) -  s \frac{T_1 T_2}{T_1+T_2} }~,
\end{align}
where $ b_s=\frac{1}{3} \times \# \,{\rm scalars}$ is the contribution to the beta function coefficient. Similar contributions would 
be included from gauge and fermion loops in a complete theory. 
Note that by an integration by parts the tensor factors have been pulled out of the integral before the integrations over proper-times. 
We can now consider the threshold contribution by subtracting the  IR logarithmic pieces with the Euclidean ``Mandelstam variable'' $s$ playing the role of RG scale. That is a gauge threshold $\Theta $ can
be defined by matching at the scale $M_s$:
 \begin{align}
\frac{16\pi^2 }{g^2_{EFT}(s) } -
\frac{16\pi^2 }{g^2_{EFT}(M_s) } &~=~ - b_s \log \left(\frac{s}{M_s^2}\right) \nonumber \\
& ~=~ \frac{16\pi^2 }{g^2(s) } -
\frac{16\pi^2 }{g^2_{\rm tree} } + \Theta ~.
\end{align}
Using the exact same approximation as for the effective potential, we can identify   
\begin{align}
\Theta  ~=~  b_s  \log \left( 4 e^{\
\gamma_E+1-\frac{\sqrt{\pi}}{2}} \right) ~.
\end{align}
\section{Multi-loops, multi-legs and saddles} 

We now turn to more intricate amplitudes, and as promised find significant simplification in many cases in the limit of high momentum.
To begin, let us consider the general structure of amplitudes obtained with our modified propagator. A general amplitude can be always reduced to a multi-dimensional integral over proper-time of the form
\begin{equation}
\mathcal{A}(\{q_i\}) \sim \int_0^\infty \prod_{i=1}^n \dd t_i e^{f \left( \{t_i + \frac{1}{t_i},q_i\} \right)}~,
\label{eq:GenAmp}
\end{equation}
where we focus primarily on the structure of the loops, and can for this discussion ignore the extra tensor structure introduced by the vertex operators. These integrals have the following properties:
\begin{align}
 \int_0^{\infty} \dd t \left( 1 - \frac{1}{t^2} \right)e^{f(t + \frac{1}{t})}  & ~=~ 0,  
\quad\implies\quad   \int_0^{\infty} \dd t e^{f(t + \frac{1}{t})} ~=~ \int_0^{1} \dd t \left( 1 +\frac{1}{t^2} \right) e^{f(t + \frac{1}{t})}~.
\end{align}
In other words, every multi-dimensional integral of this kind can be reduced to an integral on a unit hypercube. Moreover, a permutation in $\{q_i\}$ can always be countered in the integrand by a permutation of $\{ T_i \}$, rendering the amplitude fully symmetric under permutation of partial momentum sums. By extension, if we choose the particular values $q^2_i \approx s\, \forall\, i$, corresponding to having just two very hard momenta (off-shell and in the Euclidean region) while all the others are relatively soft, the integrand gains a permutation symmetry under exchange of $\{T_i\}$. Since the amplitudes are obtained from the usual ones by the replacement $T_i \rightarrow t_i + \frac{1}{t_i}$, the amplitudes in this particular limit reduce to
\begin{equation}
\mathcal{A}(\{q_i\}) \sim \int_0^\infty \prod_{i=1}^n \dd t_i e^{  - s \, f \left( \{t_i + \frac{1}{t_i}\} \right)} \sim \sum_{saddles} e^{- s \,S_{saddle}}~.
\end{equation}
In the limit $s \rightarrow \infty$ the integral is therefore dominated by its leading saddle at $t_i=1~\forall i$, which is always present since $t_i=1$ is always a solution of 
$
 \frac{\partial f}{\partial t_i} = \frac{\partial T_i}{\partial t_i} \frac{\partial f}{\partial T_i} =0. 
$ We can state the following:
\begin{quote}
 {\it The $t_i \rightarrow \frac{1}{t_i}$ proper-time symmetry implies an extremum in the action for all amplitudes in Eq.\eqref{eq:GenAmp} at $t_i=1~\forall i$. In the Euclidean region this saddle dominates the hard momentum limit. }
\end{quote}
This is very different from what happens in ordinary field theory, where the final result can have only power-like or logarithmic behaviour (coming from the introduction of a regulator scale), but is rather similar to what happens in string-theory \cite{grossmende} (although somewhat more universal, as there the positions of the leading saddles are logarithmically dependent on ratios of Mandelstam variables)\footnote{In normal field theory there is {\it some} saddle-point behaviour for the Feynman parameters (i.e. the ratios of the $T_i$'s), but by dimensions there cannot be a saddle-point for the final integration over the Schwinger parameter (i.e. the sum of the $T_i$'s).}. Note that in what follows, due to the 
exponential suppression we do not need to commit to a specific action -- as we are interested in generic behaviour, 
we will retain the possibility to have all possible $n$-point vertices. 

\noindent \underline{\it The $n$-point 1-loop amplitude}: let us now consider 
 \eqref{qmaster} in this limit. Evaluating the integral with a saddle-point approximation we find 
\begin{align}
\mathcal{A}^{(n)}_{1\ell}(\{p_i\}) &~\sim~ \delta^d(\sum p_i ) \frac{1}{(8N\pi)^{d/2}} \, e^{\frac{2}{N}\sum_{ij} q_i \cdot q_j  - 2 \sum q_i^2}\prod_{i=1}^N \sqrt{\frac{\pi}{4 \sigma_i}} + \text{perm.}~,
\end{align}
where we have introduced
\begin{align}
\sigma_{i} &~=~ \frac{1}{n^2} \sum_{kl} q_k q_l - \frac{2}{n} \sum_j q_i q_j +q_i^2 ~.
\end{align}
We can now read off the saddle-point action:
\begin{equation}
S_{saddle}^{1\ell} ~=~ -\frac{2}{N}\sum_{ij = 1}^n q_i \cdot q_j  + 2 \sum_{i=1}^n q_i^2
\end{equation}
Taking the hard momentum limit ($q^2_i = s ~\forall~ i$) we find 
\begin{equation}
\sigma_{i<n} ~=~ \frac{s}{n^2} ~~, \quad ~ \sigma_n ~=~ s \left( \frac{n-1}{n} \right)^2~,
\end{equation}
and hence 
\begin{align}
\mathcal{A}^{(n)}_{1\ell}(\{p_i\}) &~\sim~ \delta^d(\sum p_i ) \frac{\text{dim}(S_n/\mathbb{Z}_n)}{(8n\pi)^{d/2}} \, \frac{e^{-2 s \left(\frac{n-1}{n}\right)}}{s^{n/2}} \left( \frac{\pi}{4} \right)^{n/2} \frac{n^{2n}}{(n-1)^2}~.
\end{align}
As an example, the $n=2$ case has $\sigma_1 =  \sigma_2= s/4 $, and  the saddle-point approximation at large $s$ is found to be  
\begin{equation}
\mathcal{A}^{(2)}_{1\ell}(p_1,p_2)~\sim  ~\delta^d(p_1+p_2) \frac{4 \pi}{(16\pi)^{d/2}} \frac{e^{-s}}{s}~.
\end{equation}
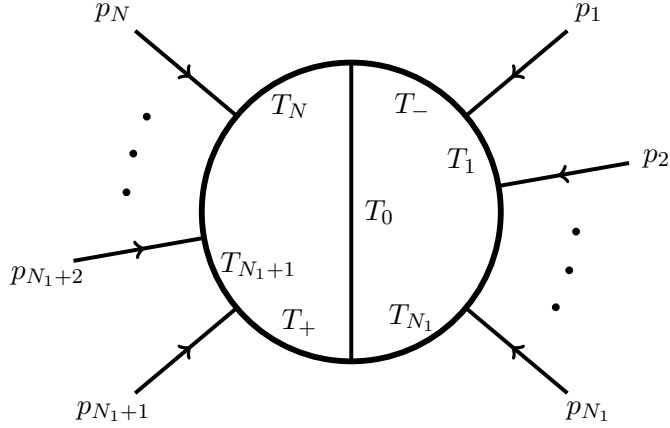
\begin{figure}[h]
\centering
	\begin{tikzpicture}[line width=1.5 pt, scale=2.5]	
		\draw[fill=black] (0,0) circle (.8cm);
		\draw[fill=white] (0,0) circle (.79cm);
		\draw[fermion] (40:1.5) -- (40:.8);
		\draw[fermion] (10:1.5) -- (10:.8);
		\draw[fermion] (-40:1.5) -- (-40:.8);
		\draw[fermion] (140:1.5) -- (140:.8);
		\draw[fermion] (-170:1.5) -- (-170:.8);
		\draw[fermion] (220:1.5) -- (220:.8);
		\draw[fermionnoarrow] (90:.8) -- (-90:.8);
		\draw[fill=black] (-5:1.2) circle (.01);
		\draw[fill=black] (-15:1.2) circle (.01);
		\draw[fill=black] (-25:1.2) circle (.01);
		\draw[fill=black] (175:1.2) circle (.01);
		\draw[fill=black] (165:1.2) circle (.01);
		\draw[fill=black] (155:1.2) circle (.01);
		\node at (40:1.65) {$p_1$};
		\node at (10:1.65) {$p_2$};
		\node at (-40:1.65) {$p_{N_1}$};
		\node at (140:1.65) {$p_{N}$};
		\node at (-168:1.65) {$p_{N_1 +2}$};
		\node at (220:1.65) {$p_{N_1+1}$};
		\node at (60:0.65) {$T_{-}$};
		\node at (25:0.65) {$T_1$};
		\node at (0:0.15) {$T_0$};
		\node at (-60:0.65) {$T_{N_1}$};
		\node at (-115:0.65) {$T_{+}$};
		\node at (-150:0.57) {$T_{N_1+1}$};
		\node at (120:0.65) {$T_{N}$};
	 \end{tikzpicture}
	 \caption{Two-loop $n$-point diagram.}
	 \label{fig:2loop}
	 \end{figure}
	 
\mbox{ } 	\\
 
\noindent \underline{\it The $n$-point 2-loop amplitude}: We now turn to the saddle-point action of the $n$-point 2-loop amplitude. The diagram shown in Fig. \ref{fig:2loop} reduces to
\begin{align}
\mathcal{A}^{(n)}_{2l,amp}(\{p_i\}) &~=~ \delta^{d}(\sum_i p_i) \int \prod \dd T_i \frac{1}{(16\pi^2 \det{M})^{d/2}} e^{- C +  V^{T} M^{-1} V } + \text{perm.}\,,
\end{align}
where we introduce the following for notational convenience:
\begin{align}
&M ~=~
\begin{pmatrix}
T_{-} + T_0 + \sum_{i=1}^{n_1} T_i & - T_0\\
- T_0  & T_{+} + T_0 + \sum_{n_1+1}^N T_i\\
\end{pmatrix} \,, \\
&V ~=~ \begin{pmatrix} 2\sum_{i=1}^{n_1} T_i q_i  ~, & 2\sum_{i=n_1 +1}^n T_i q_i  
\end{pmatrix}^T \,,\\
&C ~=~ T_{+} q_{n_1}^2 + \sum_{i=1}^n T_i (q_i^2 + m^2) \,,
\end{align}
and all Lorentz indices are suppressed. The expression for the saddle-point action is
\begin{align}
S_{saddle}^{2\ell} &~=~ 2 \left[ \sum_{i=1}^{n} q_i^2 + q_{n_1}^2 \right] \nonumber\\
& ~~\qquad - ~2 \frac{(n_2 +2) \left(\sum_{i=1}^{n_1} q_{i} \right)^2 +(n_1 +2) \left(\sum_{i=n_1+1}^{n} q_{i} \right)^2 + 2 \left( \sum_{i=1}^{n_1} q_{i} \right) \cdot \left( \sum_{i=n_1+1}^{n} q_{i} \right)}{\left[ (n_1+2)(n_2+2) -1 \right]} \,.
\end{align}
As an illustrative example, taking the hard momentum limit previously defined in the case $n_1 = n_2 = n/2$, we have
\begin{equation}
S^{2\ell}_{saddle} ~=~ 6 - \frac{8}{n+2} ~\,>\,~ 2 - \frac{2}{n} ~=~ S^{1\ell}_{saddle} \quad \forall n \,.
\end{equation}
This shows a suppression that becomes enhanced with loop order.
\begin{figure}[H]
\centering
	\begin{tikzpicture}[line width=1.5 pt, scale=2.5]	
		\draw[fill=black] (0,0) circle (.8cm);
		\draw[fill=white] (0,0) circle (.79cm);
		\draw[fermion] (0:1.3) -- (0:.8);
		\draw[fermion] (180:1.3) -- (180:.8);
		\draw[fill=black] (90:0.05) circle (.01);
		\draw[fill=black] (-90:0.15) circle (.01);
		\draw[fill=black] (-90:0.35) circle (.01);
	    	\draw[black] (0.8,0) arc (14:165:0.82cm);
		\draw[black] (0.8,0) arc (-14:-165:0.82cm);
		\draw[black] (0.8,0) arc (41:138:1.06cm);
		\node at (0:1.4) {$p_2$};
		\node at (180:1.4) {$p_1$};
	 \end{tikzpicture}
	 \label{fig:Sunset}
	 \caption{$\ell$-loop sunset diagram.}
\end{figure}
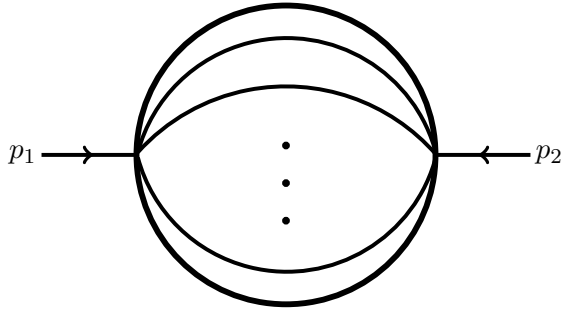

\mbox{ }

\noindent \underline{\it $\ell$-loops, 2-points -- the sunset diagram}: As a specific example of a multiloop-diagram, consider the sunset diagram with $\ell+1$ internal propagators, corresponding to an $\ell$ loop diagram. Again we will assume a trivial vertex operator for the $\ell+1$ vertexes. 
In order to write the amplitude we define the object $W^{j} _{L}$, which is the sum of all words of length $L$ that can be made with the symbols $\{ T_1, \ldots ,T_j \}$: so for example 
\begin{equation}
W^{ 5 } _4 ~=~ T_1 T_2 T_3 T_4 + T_1 T_2 T_3 T_5 + T_1 T_2 T_5 T_4 + T_1 T_5 T_3 T_4 + T_5 T_2 T_3 T_4~.
\end{equation}
After a significant manipulation, the amplitude can be written
\begin{equation}
\label{a22}
\mathcal{A}^{(2)}_{\ell}(p_1,p_2)  ~=~ \frac{\delta^4 (p_1+p_2) }{(\ell+1)!}\int \prod_{i}^{\ell+1} \dd t_i \frac{1}{\left[(4\pi)^{\ell} \sum W^{\ell+1}_{\ell}\right]^{d/2}} e^{-m^2 \sum_i^{\ell+1} T_i - p_1^2 \left[ T_{\ell+1} - T_{\ell+1}^2 \frac{W_{\ell-1}^{\ell}}{W_{\ell}^{\ell+1}} \right]}~.
\end{equation}
Taking the saddle-point approximation with $T_i \rightarrow t_i + \frac{1}{t_i}, \,\, t_i = 1 + \epsilon_i$, The action of the saddle is remarkably simple:
\begin{equation}
S_{saddle}^{sun} = -2 p_1^2 \left[ 1 - \frac{\ell}{\ell+1}\right] \,.
\end{equation}
Thus in this case, the enhanced suppression from the high number loops is compensated by the growth in leg-number of the vertices.  \\
%
%

\noindent \underline{General properties of the Amplitudes}: We end the discussion of amplitudes by conjecturing some general properties
 in the Euclidean region, that are motivated by the calculations above:
\begin{itemize}
\item At fixed loop order and fixed numbers of external legs, the leading graphs at the saddle are those with highest number of legs per vertex. 
\item At fixed loop order and fixed numbers of legs per vertex, the leading graphs at the saddle are those with the fewest vertex insertions.
\item At fixed numbers of external legs and fixed numbers of legs per vertex, the leading graphs at the saddle are those with the lowest  loop order. If a suitable tree vertex is present, this is the leading one.
\end{itemize}

\begin{figure}[H]
\begin{center}
\begin{tikzpicture}[scale=2.4]
\begin{scope}[very thick,decoration={
    markings,
    mark=at position 0.35 with {\arrow{<}}}
    ] 

\def\gap{0.5mm}
\def\bigradius{1.3cm}
\def\smallradius{0.5mm}

\node[circle] at (0.8*\bigradius,0.8*\bigradius) {\huge $A$};
\node[circle] at (-0.8*\bigradius,0.8*\bigradius) {\huge $B$};
\node[circle] at (0.8*\bigradius,-0.8*\bigradius) {\huge $D$};
\node[circle] at (-0.8*\bigradius,-0.8*\bigradius) {\huge $C$};
\node[circle] at (+2.1mm,6mm) {\huge $ \gamma_-$};
\node[red] at (-2.8mm,9mm) {\huge $\gamma_F$};
\node[circle] at (-2.0mm,-6mm) {\huge $\gamma_+$};

\draw[dashed, very thick, postaction={decorate}] (\bigradius,0.5mm) arc [start angle=2, end angle=88, radius=\bigradius]  ;
\draw[dashed, very thick, postaction={decorate}] (-0.5mm,\bigradius) arc [start angle=92, end angle=178, radius=\bigradius];
\draw[dashed, very thick, postaction={decorate}] (\bigradius,-0.5mm) arc [start angle=358, end angle=272, radius=\bigradius];
\draw[dashed, very thick, postaction={decorate}] (-0.5mm,-\bigradius) arc [start angle=268, end angle=182, radius=\bigradius]  ;
\draw[very thick,postaction={decorate}]
      (\gap,-\bigradius) -> (\gap,-0.2*\bigradius) arc (0:180:\smallradius) (-\gap,-0.2*\bigradius) --  (-\gap,-\bigradius) ;
\draw[very thick,postaction={decorate}]
      (\gap,\bigradius) -> (\gap,0.2*\bigradius) arc (0:-180:\smallradius) (-\gap,0.2*\bigradius) --  (-\gap,\bigradius) ;      
\draw [very thin] (-1.5*\bigradius,0) -- (1.5*\bigradius,0) node [above left]  {\huge $\Re(p_0)$ \hspace{-0.4cm} };
    \draw [very thin] (0,-1.25*\bigradius) -- (0,1.25*\bigradius) node [below left = -3pt] {\huge $\Im(p_0)$};
\draw[xshift = 0.01mm, snake,gray!50] (0, 0.2*\bigradius) -- (0, 1.25*\bigradius);
\draw[xshift = 0.01mm, snake,gray!50] (0, -0.2*\bigradius) -- (0, -1.25*\bigradius);
\draw[fill] (0,0.2*\bigradius) circle [radius=0.2mm];
\draw[fill] (0,-0.2*\bigradius) circle [radius=0.2mm];
\end{scope}

\begin{scope}[very thick,decoration={
    markings,
    mark=at position 0.47 with {\arrow{>}}}
    ] 
\def\gap{0.5mm}
\def\bigradius{1.3cm}
\def\smallradius{0.5mm}  
\clip(0,0) circle(\bigradius);
\draw[blue,very thick,postaction={decorate}]
      (-\bigradius,0.1*\gap) -- (\bigradius,0.1*\gap) ;
      (-\bigradius,-\gap) -- (\bigradius,-\gap) ;
      \node[blue] at (-0.7*\bigradius,3*\gap ) {\huge $\gamma_E$};


\draw[red, ultra thick,postaction={decorate}, domain=- 0.1:0.1] plot (\x, { -0.1*tan(89/0.1* \x) });

\end{scope}
\end{tikzpicture}
\caption{Contour for would-be ``Wick rotation'' in the presence of branch-cuts, where $p_0$ is real for the Euclidean case. $\gamma_E$ is the contour for the Euclidean propagator, while $\gamma_F$ is the contour for the ($+ i \epsilon $)  Feynman propagator in Minkowski space. At infinity, the integrals for any propagator defined as a function of $p^2$, obey $B=D=-A^*$ and  $C=A$. Equating the Euclidean propagator $\Delta$ with the semi-circular integrals taken above and below the real axis gives $\gamma_- = \gamma_+$. \label{wick}}
\end{center}
\end{figure}
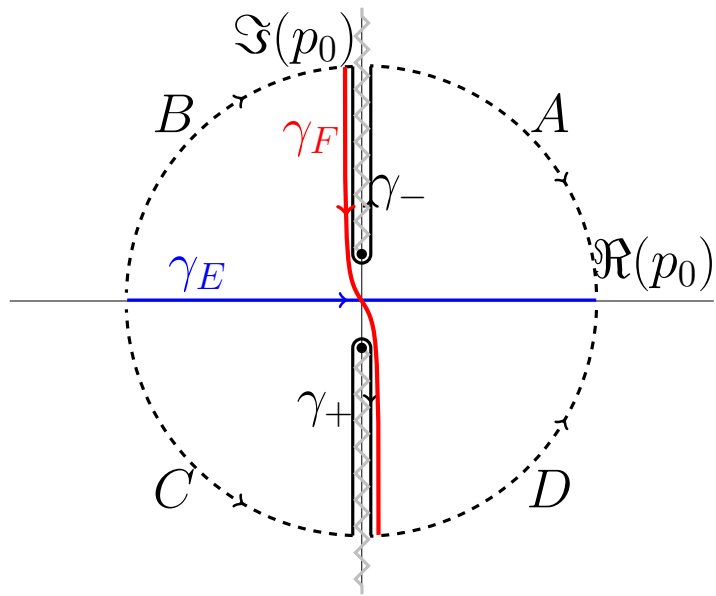

\section{Remarks on the picture in Minkowski space}

Despite the simplicity of performing amplitudes in the Euclidean picture, the  passage from Euclidean to Minkowski space in infinite derivative field theories is a delicate issue, and requires further comment. As we shall see the worldline procedure casts some light on this question.  

First we give a summary of the issues as they are typically presented. We will consider the propagator of the special  time-inversion symmetric case 
\be
\Delta(p^2) ~=~ 2  K_1 (2 (p^2+m^2)) ~,
\ee
and will attempt to analytically continue the Euclidean propagator to Minkowski coordinate-space. 
That is we wish to use a Wick rotation to evaluate    
\begin{align}
\Delta_F(x,y) &~=~ \intp{p} e^{ip(x-y)}  2 i  K_1 (2 (p^2 + m^2 - i\epsilon))  ~,
\end{align}
where we use the mostly plus signature, so that $p^2=\vec{p}^2-p_0^2$, and where the $i\epsilon$ prescription is determined by the 
limit in \eqref{delta}.
We start by splitting the momentum integral into the 3-vector $\vec{p}$, and the time component which we wish to Wick rotate, $p_0$. It is  convenient to display the contours required for the Wick rotation to pass to Minkowski space as in 
Figure \ref{wick}.  The Bessel function possesses a branch-cut along the negative $p^2$ axis, from the pole at $p^2=-m^2$, which translates into branch-cuts in the $p_0$-plane.  Specifically, inside the 3-momentum integral there are the usual poles at $p_0 = \pm i E_p$ where $E_p=\sqrt{\vec{p}^2 +m^2}$, and the branch-cuts go from here to $\pm i\infty$.   Thus the Euclidean propagator $\Delta$ corresponds to the contour $\gamma_E$, while the standard 90-degree rotation clockwise to the Feynman propagator $\Delta_F$  in Minkowski space corresponds to $\gamma_F$. An obstacle now arises, because in order to Wick rotate between the two, the contours at infinity, $A = C$, should go to zero, but they do not. Instead they begin to diverge once the Wick rotation angle goes beyond $\pi/4$, due to the exponential form-factor, which goes as $e^{-2(p^2+m^2)}$ at large radius. Therefore it does not seem to be possible to define a Feynman $\Delta_F$ that is related by a simple Wick rotation to the Euclidean $\Delta$.
The typical proposal for dealing with this issue in infinite derivative field theories is to 
proceed with the calculation in Euclidean formalism, and then to pass to Minkowski space only at the end 
of the calculation once the amplitude has been determined (see for example  
\cite{Buoninfante:2018mre} and \cite{Pius:2016jsl} for further discussion on this point).

The present worldline prescription makes the situation a little clearer. The entire calculation can be formulated using the worldline theory, 
with the attendant worldline Green's function, and the vertex operators, and without reference to integration over 
internal momenta. The only momenta appearing in the calculation are in the physical Mandelstam variables of the external states, and therefore the
UV finiteness of the theory does not rely on the exponential suppression of propagators.  Indeed if we consider the theory with worldline inversion symmetry for example, the UV region of any integral at $t\rightarrow 0$ is equivalent to the IR region at $t\rightarrow\infty$, so 
there can {\it only} be IR  divergences in the amplitudes. As in the vacuum polarisation calculation of \eqref{thresh}, a consistent procedure is to then compute the amplitude in a Euclidean region of phase space (where $s>0$), and analytically continue 
to time-like regions. This for example will pick up the imaginary contributions in the amplitudes that one expects from the optical theorem, 
when states in the loop can go on-shell. Of course the integration over Schwinger proper-time in such regions would diverge, but 
we can be confident that these divergences are just logarithmic IR ones. 

Thus the behaviour when $|s|\ll 1 $ is well understood for either sign of $s$, as being that of a consistent effective field theory with a finite UV completion. However 
this  does not address the behaviour of the amplitudes when the external momenta themselves become large and time-like, when  they appear to blow-up as $e^{|s|}$, threatening unitarity. (Note that when $s<0$ there is no saddle but the integral would still appear to grow exponentially).  

In this regime, one can argue in a more heuristic way that the amplitudes are still exponentially suppressed, by re-organising the perturbative expansion, such that all internal propagators are replaced by fully dressed ones (see for example \cite{Buoninfante:2018mre}). In the 
present case this consists of the replacement (we neglect masses in this limit):
\begin{equation}
2K_1(2p^2) \rightarrow \frac{ 2K_1(2p^2) }{ 1 - 2\Sigma(p^2) K_1(2p^2) }~,
\label{eq:fullProp}
\end{equation}
where $\Sigma$ is the 1PI amplitude. In the space-like region where everything is perturbative, expanding this propagator shows it to be a resummation of bubbles. Analytically continuing this expression to the hard time-like region, we find exponentially suppressed propagators. As an example, consider highly energetic $2 \rightarrow 2$ scattering in the $s$-channel (where $s=(p_1+p_2)^2<0$ in mostly plus signature):
\begin{equation}
\mathcal{A}_{2\rightarrow 2,s}(s) ~\sim~ 2 K_1(2 s)~ =~ - 2 K_1(-2s) \pm 2i \pi I_1 (-2s) ~\sim~  i\, e^{-2s}~.
\end{equation}
Note that the amplitude for negative $s$ becomes entirely imaginary, with the dominant contribution coming from the branch-cut, which 
can thus be thought of as representing a growing continuum of unstable states, which do not however appear in the asymptotic Hilbert space. The picture then becomes reminiscent of the regularisation
observed in \cite{Khoze:2017tjt}\footnote{The approaches are quite different however: in \cite{Khoze:2017tjt} amplitude growth in the IR is the cure for UV divergences, whereas in the present context UV/IR mixing (which is a typical feature of regularising UV divergences) is a fundamental principle imposed on the worldline, and exponential suppression in the propagators is the outcome.},
with the full propagators  in  (\ref{eq:fullProp}) 
taming this apparent growth in the amplitude. In the present case of a quartic scalar theory the first contribution to the IPI amplitude $\Sigma$ to blow up is the two-loop sunset diagram, as in \eqref{a22}, which also scales as $e^{-2s}$.

\section{Conclusions}

This paper has proposed an extremely simple worldline formalism for UV completing particle theories. 
The procedure takes inspiration from the formulation and behaviour of first quantised string-theory. 
The result can be directly related to a novel kind of infinite-derivative ghost-free field theory, although 
using an {\it ab initio} worldline framework makes calculation very straightforward.
An attractive choice is then to mimic the modular invariance of string-theory by imposing a worldline inversion symmetry, because this guarantees that all perturbative divergences can be interpreted as infra-red ones. 
It also leads to simplification in amplitudes, whose UV sensitive contributions 
are dominated by saddle-points. While the discussion focussed on scalar fields, the worldline 
definition can easily be extended to particles with spin. 

While theories of minimal length have appeared in the literature before (see \cite{Hossenfelder:2012jw} for a review), these are 
typically predicated on the notion of coordinate-space duality. The proposal here is based on a modular-invariance like symmetry 
imposed on the worldline itself, and yields completely different results, indeed giving a lower bound on the proper-time. 

At first sight our procedure seems to be a quite brutalist insertion of 
string-theory features into the worldline formalism of particle-theory, and yet we are unable to find anything obviously  
wrong with the result. It seems neither more nor less consistent than other non-local field theories that have been studied in the literature, and has very significant advantages.  Nevertheless the physical meaning remains intriguing: the procedure does not for example appear to correspond to a limit of string-theory. One could simply regard it as a UV completion in its own right, but it would also be of great interest to find a microscopic derivation of such theories.  \\

\vspace{0.3cm} 
\noindent {\bf Acknowledgements:}
SAA would like to thank the GGI for hospitality during which some of the ideas presented here were first road-tested, Valya Khoze for conversations, and David Skinner and Ashok Sen for comments and questions that led to clarifications in the text.   NAD would like to thank the IPPP for kind hospitality during the whole duration of this project. The CP$^3$-Origins centre is partially funded by the Danish National Research Foundation, grant number DNRF:90. This work was made possible by EU directive 2004/38/EC on the right to free movement.

\end{document}